\documentclass[pra,twocolumn,showpacs,amsmath,amssymb,aps,superscriptaddress]{revtex4-1}

\usepackage{graphicx}% Include figure files
\usepackage{dcolumn}% Align table columns on decimal point
\usepackage{bm}% bold math
\usepackage{txfonts}
\usepackage{amsmath}
\usepackage{epsf}
\usepackage{epsfig}
\usepackage{amssymb}
\usepackage{color}
\usepackage{multirow}
\usepackage{pst-3d}
\usepackage{rotating}

\begin{document}

\title{Simulating open quantum systems by applying SU(4) to quantum master
    equations}

\author{Minghui Xu} \author{D. A. Tieri} \author{M. J. Holland}
\affiliation{JILA, National Institute of Standards and Technology and
  Department of Physics, University of Colorado, Boulder, Colorado
  80309-0440, USA}

\date{\today}%

\begin{abstract}
  We show that open quantum systems of two-level atoms symmetrically
  coupled to a single-mode photon field can be efficiently simulated
  by applying a {\em SU}(4) group theory to quantum master
  equations. This is important since many foundational examples in
  quantum optics fall into this class. We demonstrate the method by
  finding exact solutions for many-atom open quantum systems such as
  lasing and steady state superradiance.
\end{abstract}


%Use showkeys class option if keyword
                              %display desired
\maketitle

\section{Introduction}
Most physical situations to which quantum mechanics is applied are
open. The open nature is necessary to treat basic irreversible
processes such as energy transfer with a heat bath, particle exchange
with a reservoir, and quantum measurements. Open quantum systems can
be treated under the Born and Markov approximations by the quantum
master equation in the Lindblad form~\cite{Lindblad76}, which has been
applied across many fields of physics, including quantum optics and
quantum information science~\cite{Gardiner04,Carmichael}, atomic and
molecular physics~\cite{Bollinger10}, solid state
physics~\cite{Blais07}, and optomechanics~\cite{Teufel11}.

In general, for all but the smallest system sizes, exact analytic
solutions to the quantum master equation are intractable. Various approximation
methods have been introduced, {\em e.g.} perturbation theories~\cite{Cirac12}, mean-field
approaches~\cite{Zoller10,nagy11}, cummulant expansions~\cite{Meiser09,Meiser10}, 
linear response theories~\cite{Lukin13} and $c$-number Langevin equations~\cite{Scully90, Fabre93}.  
However, it is often necessary to benchmark approximate methods with exact numerical solutions.
Existing numerical simulation approaches, such as the quantum Monte
Carlo method~\cite{Knight98}, scale exponentially with
 the underlying dimensionality of the Hilbert space. Therefore, treating 
any appreciable system size is extremely difficult.

Here we present a novel group-theoretic approach to find an efficient
solution of the quantum master equation, which reduces the
exponential scaling of the problem to cubic. Even though we focus on an
important class of quantum optical systems, the methods we present
could be more generally applied. 
We consider the symmetric coupling of
a single-mode cavity field to an ensemble of $N$ two-level atoms
(analogous to pseudo-spin-1/2 systems or qubits). The Hamiltonian that
describes this situation in the interaction picture is given by
\begin{equation}
  H=\frac{\hbar\Delta}{2}\sum_{j=1}^N\sigma_j^{(3)}
  +\hbar \Omega\sum_{j=1}^{N}(a^\dagger\sigma_j^-+a\sigma_j^+)\,,
\label{eq:hamiltonian}
\end{equation}
where the first term is the free energy, with $\Delta$ being the
detuning of the light field from the atomic transition, and the second
term is the reversible atom-field coupling with strength~$\Omega$. The
photon annihilation operator is $a$, and $\sigma_j^{(3)}$ and
$\sigma_j^+=(\sigma_j^-)^\dagger$ are Pauli operators for the $j$th
spin-component. In the presence of decoherence, the full quantum
evolution is described by the quantum master equation for the reduced
density operator $\rho$:
\begin{eqnarray}\label{eq1}
\dot{\rho} &=& \mathcal{L}\rho=
  \frac{1}{i\hbar}[H,\rho]+\kappa\mathcal{D}[a]\rho\nonumber\\
  &&+ \sum_{j=1}^N\left(\gamma\mathcal{D}[\sigma_j^-]
    +w\mathcal{D}[\sigma_j^+]+\frac{1}{2T_2}
    \mathcal{D}[\sigma_j^3]\right)\rho\,,
\end{eqnarray}
where $\mathcal{D}[\hat{O}]\rho=(2\hat{O}\rho
\hat{O}^\dagger-\hat{O}^\dagger \hat{O}\rho-\rho \hat{O}^\dagger
\hat{O})/2$ denotes the Lindblad superoperator. We have introduced the
decay rate $\kappa$ for the cavity, and population relaxation rates
for the spin components $\gamma,w$ (for decay and pumping
respectively) and dephasing rate $1/(2T_2)$.

%It should be
%emphasized that the fully-symmetrical Dicke states do not suffice as a
%reduced basis since basis vectors with mixed symmetry are required.

\section{Applying {\em  SU}(4) to the quantum master equation}
Recently, it was pointed out that it is preferable to work in
Liouville space rather than in Hilbert space since the Lindblad operators are invariant under {\em
  SU}(4) transformations~\cite{Hartmann12}. This observation allows
one to express all of the Lindblad operators in terms of generators of
the {\em SU}(4) group. For this purpose, 18 superoperators
$\mathcal{O}_+$, $\mathcal{O}_-$ and $\mathcal{O}_3$ where
$\mathcal{O}\in\left\{\mathcal{Q},\Sigma,\mathcal{M},\mathcal{N},
  \mathcal{U},\mathcal{V}\right\}$ are defined
\begin{equation}\label{super}
\begin{split}
  \mathcal{Q}_{\pm}\rho :=
  \sum_{j=1}^N\sigma_j^{\pm}\rho\sigma_j^{\mp}&,\;\;\;\mathcal{Q}_3\rho
  :=
  \frac{1}{4}\sum_{j=1}^N\left(\sigma_j^3\rho+\rho\sigma_j^3\right)\\
  \Sigma_{\pm}\rho :=
  \sum_{j=1}^N\sigma_j^{\pm}\rho\sigma_j^{\pm}&,\;\;\;\Sigma_3\rho :=
  \frac{1}{4}\sum_{j=1}^N\left(\sigma_j^3\rho-\rho\sigma_j^3\right)\\
  \mathcal{M}_{\pm}\rho :=
  \sum_{j=1}^N\sigma_j^{\pm}\rho\frac{1+\sigma_j^3}{2}&,\;\;\;\mathcal{M}_3\rho
  :=
  \frac{1}{2}\sum_{j=1}^N\sigma_j^3\rho\frac{1+\sigma_j^3}{2}\\
  \mathcal{N}_{\pm}\rho :=
  \sum_{j=1}^N\sigma_j^{\pm}\rho\frac{1-\sigma_j^3}{2}&,\;\;\;\mathcal{N}_3\rho
  :=
  \frac{1}{2}\sum_{j=1}^N\sigma_j^3\rho\frac{1-\sigma_j^3}{2}\\
  \mathcal{U}_{\pm}\rho :=
  \sum_{j=1}^N\frac{1+\sigma_j^3}{2}\rho\sigma_j^{\mp}&,\;\;\;\mathcal{U}_3\rho
  :=
  \frac{1}{2}\sum_{j=1}^N\frac{1+\sigma_j^3}{2}\rho\sigma_j^3\\
  \mathcal{V}_{\pm}\rho :=
  \sum_{j=1}^N\frac{1-\sigma_j^3}{2}\rho\sigma_j^{\mp}&,\;\;\;\mathcal{V}_3\rho
  := \frac{1}{2}\sum_{j=1}^N\frac{1-\sigma_j^3}{2}\rho\sigma_j^3.
\end{split}
\end{equation}
Although this list, Eq.~(\ref{super}), contains 18 operator
definitions, only 15 of them are independent (it is possible to write
$\mathcal{N}_3$, $\mathcal{U}_3$, $\mathcal{V}_3$ in terms of the
others). One can also demonstrate that the 15 remaining superoperators are
linear combinations of the familiar Gell-Mann matrices that are the
generators of the {\em SU}(4) group,
$\lambda_1,...,\lambda_{15}$~ (see Appendix \ref{app1}).

As a consequence, it is possible to construct a reduced basis for the
density operator using a multiplet of the {\em SU}(4)
group. Transcribing notation from the four-flavor quark model---a
model with the same symmetry structure---the fundamental
representation is given by $u=|1\rangle\langle1|$,
$d=|0\rangle\langle0|$, $s=|1\rangle\langle0|$, and
$c=|0\rangle\langle1|$ (up, down, strange, and charm).  Since the
symmetry type of the basis is preserved under the action of the {\em
  SU}(4) generators~\cite{note1}, this leads to a tremendous reduction
of the number of required basis states needed to provide an exact
solution of the master equation.

For the fully symmetric case, the basis is:
\begin{equation}
  P_{q,q_3,\sigma_3}=\mathcal{S}(u^\alpha d^\beta
  s^\gamma c^\delta),
\end{equation}
where $\mathcal{S}$ denotes the symmetrizer and
$\alpha+\beta+\gamma+\delta=N$. Note that only basis states with
$\gamma=\delta=0$ have non-vanishing trace. The three quantum numbers
$q,q_3$ and $\sigma_3$ have ranges $q=0,1/2,...,N/2$,
$q_3=-q,-q+1,...,q$ and $\sigma_3=q-N/2, q-N/2+1,...,N/2-q$, resulting
in the dimensionality of the basis $(N+1)(N+2)(N+3)/6$, {\em i.e.\/}
of order $N^3$. This tremendous reduction should be compared with the
full dimensionality of the Liouville space given by~$4^N$.

In this paper, we apply the {\em SU}(4) group theory to find exact
solutions to the quantum master equation in general form. We show how
to calculate the various basic observables of interest.  We
demonstrate that the density matrix in the {\em SU}(4) basis
representation can be precisely mapped to the collective
spin-angular-momentum representation $|S,M\rangle$ in Hilbert space,
which enables us to efficiently diagonalize the density matrix. This
allows us to provide complete information about the system, including
functional properties of the density operator such as the purity and
von Neumann entropy.

In order to solve Eq.~(\ref{eq1}), we expand the density matrix as
\begin{equation}\label{ex}
  \rho=\sum_{q,q_3,\sigma_3,m,n} C_{q,q_3,\sigma_3}^{m,n}
  P_{q,q_3,\sigma_3}\bigl|m\bigr>\bigl<n\bigr|\,,
\end{equation}
where $C_{q,q_3,\sigma_3}^{m,n}$ are complex coefficients, and
$|n\rangle$ is the photon Fock state.  The Lindblad operators can be
written compactly:
\begin{eqnarray}\label{liv}
  \sum_{j=1}^N\mathcal{D}[\sigma_j^{\pm}]&=&-\frac{N}{2}\pm
  \mathcal{Q}_3+\mathcal{Q}_{\pm}\,,\nonumber\\
  \sum_{j=1}^N\mathcal{D}[\sigma_j^{(3)}]&=&4\mathcal{M}_3-2
  \mathcal{Q}_3-2\Sigma_3-N\,.
\end{eqnarray}
The completeness of $\mathcal{O}_{+,-,3}$ and $a$ implies that an
arbitrary Hamiltonian can be expressed by them, {\em e.g.}  from
Eq.~(\ref{eq:hamiltonian}),
\begin{eqnarray}\label{ham}
  \frac{1}{i\hbar}[H,\rho]&=&-2i\Delta\Sigma_3\rho-i\Omega
  \left[a(\mathcal{M}_++\mathcal{N}_+)\rho+a^\dagger
    (\mathcal{M}_-+\mathcal{N}_-)\rho\right]\nonumber\\
  &&\quad{}+i\Omega\left[(\mathcal{U}_++\mathcal{V}_+)\rho a^\dagger
    +(\mathcal{U}_-+\mathcal{V}_-)\rho a\right]\,.
\end{eqnarray}
Combining Eqs.~(\ref{liv}) and (\ref{ham}) with the action rules of
the {\em SU}(4) and photon operators on the basis states (see
Appendix \ref{app2}) gives a closed solution of Eq.~(\ref{eq1}). In
general, this can be solved analytically or numerically with standard
methods.
\section{Observables}
Having established the procedure for determining the time evolution of
$\rho$, it is now important to describe how to calculate physical
observables. We begin with the trace given by:
\begin{equation}
\mathrm{Tr}[\rho]=\sum_{m,q3}C_{N/2,q3,0}^{m,m}=1\,,
\label{eq:trace}
\end{equation}
which is an invariant during evolution to represent probability
conservation. Average values $\bigl<a\bigr>$ and $\bigl<a^\dagger
a\bigr>$ are found analogously. For the spin-operators, we provide the
following examples up to quadratic order:
\begin{equation}\label{ob}
\begin{split}
  &\langle\sigma_j^{(3)}\rangle=2\mathrm{Tr}[\mathcal{Q}_3\rho]/N,\\
  &\langle\sigma_j^{(3)}\sigma_k^{(3)}\rangle=(4\mathrm{Tr}
  [(\mathcal{Q}_3^2-\Sigma_3^2)\rho]-N)/[N(N-1)],\\
  &\langle\sigma_j^{\pm}\rangle=\mathrm{Tr}[(\mathcal{M}_{\pm}
  +\mathcal{N}_{\pm})\rho]/N,\\
  &\langle\sigma_j^+\sigma_k^-\rangle=\mathrm{Tr}
  [\mathcal{V}_-(\mathcal{M}_-+\mathcal{N}_-)\rho-
  \mathcal{Q}_-\rho]/[N(N-1)],
\end{split}
\end{equation}
where $j\ne k$. %We do not provide $j\!=\!k$ quantities since they
%follow easily from the basic algebra of the Pauli spin-operators.

For coherence properties it is necessary to calculate products of
operators evaluated at different times. Of particular interest are the
first-order and second-order correlations, which can be found by
applying the quantum regression theorem:
\begin{eqnarray}\label{quantumre}
  \langle\hat{O}_1(t+\tau)\hat{O}_2(t)\rangle&=&\mathrm{Tr}
  \left[\hat{O}_1e^{\mathcal{L}\tau}[\hat{O}_2\rho(t)]\right],
  \nonumber\\ \hspace*{-1pc}
  \langle\hat{O}_1(t)\hat{O}_1(t\!+\!\tau)\hat{O}_2(t\!+\!\tau)
  \hat{O}_2(t)\rangle
  &=&\mathrm{Tr}\left[\hat{O}_2e^{\mathcal{L}\tau}
    [\hat{O}_2\rho(t)\hat{O}_1]\hat{O}_1\right],
\end{eqnarray}
where $e^{\mathcal{L}\tau}[\rho]$ is the time propagation from
Eq.~(\ref{eq1}) starting with the initial density matrix $\rho$. For
example, in order to obtain the first-order correlation of $\hat{O}_1$
and $\hat{O}_2$, one takes $\hat{O}_2\rho(t)$ as an initial condition,
time evolves it for $\tau$ according to Eq.~(\ref{eq1}), applies
$\hat{O}_1$, and computes the trace. A similar procedure follows for
the second-order correlation. In this way, field quantities, $\langle
a^{\dagger}(t+\tau)a(t)\rangle$ and $\langle a^{\dagger}(t)
a^{\dagger}(t+\tau)a(t+\tau)a(t)\rangle$ are directly calculated. For
spin-coherence, the required expressions are:
\begin{eqnarray}
  &&\sum_{j,k=1}^{N}\langle \sigma_j^+(t+\tau)\sigma_k^-(t)\rangle
  =\mathrm{Tr}\left[(\mathcal{M}_++\mathcal{N}_+)e^{\mathcal{L}\tau}
    [(\mathcal{M}_-+\mathcal{N}_-)\rho(t)]\right],\nonumber\\
  &&\sum_{j,j',k,k'=1}^{N}\langle \sigma_j^+(t)\sigma_{j'}^+
  (t+\tau)\sigma_k^-(t+\tau)\sigma_{k'}^-(t)\rangle=\nonumber\\
  &&\qquad\mathrm{Tr}\left[\mathcal{V}_-(\mathcal{M}_-+\mathcal{N}_-)
    e^{\mathcal{L}\tau}[\mathcal{V}_-(\mathcal{M}_-+\mathcal{N}_-)\rho(t)]\right]\,.
\end{eqnarray}
\section{Transform to the $|S,M\rangle\langle S,M'|$ representation}
Although at this point we have provided a theoretical framework that
is complete and provides exact and efficient solutions to the general
quantum master equation, it is often inconvenient to work in the
$P_{q,q_3,\sigma_3}$ representation of the density operator. For
example, it can be a nontrivial procedure to characterize the
many-body spin-state in this representation by quantifying the degree
of entanglement, which is derived from a functional ({\em i.e.}~${\rm
  Tr}[\rho\log(\rho)]$). For this reason, we illustrate now the
procedure for efficiently projecting the density operator from the
{\em SU}(4) basis representation onto the usual representation of
density matrices formed from the Hilbert space basis vectors. These
Hilbert space basis vectors are specified by the angular momentum
eigenket $|S,M\rangle$, where $S=N/2,N/2-1,...,(1/2 \,\mathrm{or}\,0)$
is the total spin and $M=-S,-S\!+\!1,\ldots,S$ is the
spin-projection. Note that $S$ also labels the symmetry of the states,
e.g. $S=N/2$ corresponds to the fully symmetrical Dicke states.

In order to illustrate how this projection is done, it is instructive
for us to first examine explicitly the $N\!=\!2$ case where the
Hilbert space is 4~dimensional. Two spins form a symmetric triplet
state and an antisymmetric singlet state, corresponding to total spin
$S=1$ and $S=0$ respectively. In this case, the complete density
matrix from Eq.~(\ref{ex}) for given ${m,n}$ is
\begin{equation}\nonumber
  \bordermatrix{
    &\langle1,1|&\langle1,0|&\langle1,-1|&\langle0,0|\cr%
    |1,1\rangle&C_{1,1,0}^{m,n}&\frac{C_{1/2,1/2,1/2}^{m,n}}{\sqrt{2}}
    &C_{0,0,1}^{m,n}&0\cr
    |1,0\rangle&\frac{C_{1/2,1/2,-1/2}^{m,n} }{\sqrt{2}}
    &\frac{C_{1,0,0}^{m,n}+C_{0,0,0}^{m,n}}{2}
    &\frac{C_{1/2,-1/2,1/2}^{m,n}}{\sqrt{2}}&0\cr
    |1,-1\rangle&C_{0,0,-1}^{m,n}&\frac{C_{1/2,-1/2,-1/2}^{m,n}}{\sqrt{2}}
    &C_{1,-1,0}^{m,n} &0\cr
    |0,0\rangle&0&0&0&\frac{C_{1,0,0}^{m,n}-C_{0,0,0}^{m,n}}{2}
  }.
\end{equation}
Notice that the resulting matrix is block diagonal in the $S=1$ and
$S=0$ subspaces (a $3\times3$ block and a $1\times1$ block). In
addition, the complex coefficients contributing to the matrix element
for $|S,M\rangle\langle S,M'|$ all satisfy $q_3+\sigma_3=M$ and
$q_3-\sigma_3=M'$.  Finally, the trace is simply
$\sum_{q_3=-1}^1C_{1,q3,0}^{m,n}=1$.

\begin{figure}[h]
\centerline{\includegraphics[width=0.9\linewidth, angle=0]{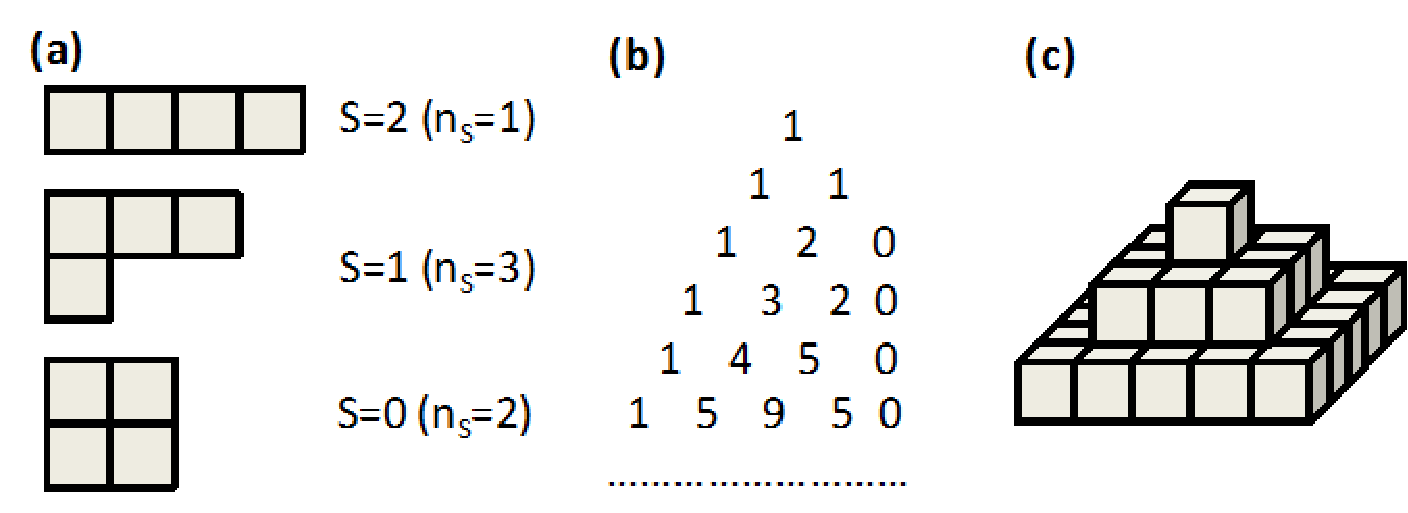}}
\caption{\label{Fig1} (a) Young tableau for determining the
  irreducible representations contained in the direct product
  representation of the {\em SU}(2) group for $N=4$. The dimension for
  $S=2,1,0$ is $5,3,1$ respectively, and $n_s=1,3,2$, so the total
  Hilbert space dimension is $5\times1+3\times3+1\times2=2^4$ as
  expected. (b) ``Pascal's triangle" to evaluate $n_S$ in an iterative
  way for any $N$ (the case considered in (a) is the fourth row down
  from the top). (c) Pyramid representation of the density operator in
  the $|S,M\rangle$ representation for $N=4$. Each layer of the pyramid
  represents the block matrix for each $S$.}
\end{figure}

These results can be systematically extended to higher $N$. For any
$N$, the density matrix is block diagonal in $S$, with each block
given by
\begin{equation}
  \rho_S^{m,n}=\sum_{M,M'}
  D_{S,M,M'}^{m,n}|S,M\rangle\langle S,M'|,
\end{equation}
where $D_{S,M,M'}$ are density matrix elements for the symmetry
type~$S$. There are $n_S$ ways for $N$ spins to construct the basis
for each~$S$, so that $\sum_{S}(2S+1)n_S=2^N$, {\em i.e.\/} the
Hilbert space dimension~\cite{Gilmore72}.  To find $n_S$, we note that $|S,M\rangle$
forms a basis of the $(2S+1)$-dimensional irreducible representation
of the {\em SU}(2) group. Determining $n_S$ is accomplished with the
help of the Young tableau of the {\em SU}(2) group, where one can
obtain the number of equivalent representations
iteratively. Fig.~\ref{Fig1}(a) shows the Young tableau for the $N=4$
case. A corresponding tabular method for evaluating $n_S$ for any $N$
is shown in Fig.~\ref{Fig1}(b), which contains about one half of
Pascal's triangle.

With this in mind, one can now derive a systematic algorithm for
obtaining density matrix elements $D_{S,M,M'}^{m,n}$ given {\em SU}(4)
expansion coefficients $C_{q,q_3,\sigma_3}^{m,n}$. The procedure is
outlined as follows. For each layer of the
pyramid~[cf. Fig.~\ref{Fig1}(c)], one may start with a corner element
($M$ and $M'$ maximal) and fill out the matrix by successive
application of the angular momentum lowering operator
$\hat{J}_-=\sum_{j=1}^N\sigma_j^-$ (noting that
$\rho\hat{J}_-=(\mathcal{U}_++\mathcal{V}_+)\rho$) to recursively fill
out each row, and $\hat{J}_-\rho$ (or hermiticity of $\rho$) to fill
out each column. The layers are filled upwards from the base, starting
with $D_{N/2,N/2,N/2}^{m,n}=C_{N/2,N/2,0}^{m,n}$ as the corner element
of the lowest layer, and finding the corner element of higher layers
by Gaussian elimination from the trace constraint
Eq.~(\ref{eq:trace}). In Appendix \ref{app3}, we demonstrate
explicit application to 3 atoms, with extrapolation to higher $N$
straightforward.

Being able to express the density operator in the $|S,M\rangle$
representation makes easy the calculation of functionals, such as the
purity $\mathrm{Tr}[\rho^2]$, or the von Neumann entropy
\begin{equation}
  S=-\mathrm{Tr}(\rho\ln\rho)=-\sum_{j}
  \lambda_j\ln\lambda_j,
\end{equation}
where $\lambda_j$ are eigenvalues of $\rho$. The point is that,
because the density matrix is block diagonal in the $|S,M\rangle$
representation, we do not need to diagonalize the whole density
matrix, which would be a daunting task. Instead, we only need to
diagonalize a series of $\lfloor N/2\rfloor+1$ blocks of dimension
$2S+1$.
\section{Application to Lasing}
In the following, we demonstrate the method by solving many-atom open
quantum systems such as lasing and steady state superradiance.  We
show the capability for finding exact solutions of large systems and
are able to obtain full information about both the transient and
steady-state density matrix.

First, let us consider a single-mode laser consisting of an ensemble
of two-level atoms coupled to an optical cavity, which can be modeled
by the general quantum master equation
Eq.~(\ref{eq1})~\cite{Carmichael}. In this model we will ignore $T_2$
dephasing for simplicity. The laser system is difficult to solve
without approximation since it involves both many atoms and large
numbers of photons when above threshold. Therefore, it constitutes an
interesting test-case to illustrate the capability of the {\em SU}(4)
approach.

\begin{figure}[h]
  \centerline{\includegraphics[width=0.75\linewidth, angle=0]{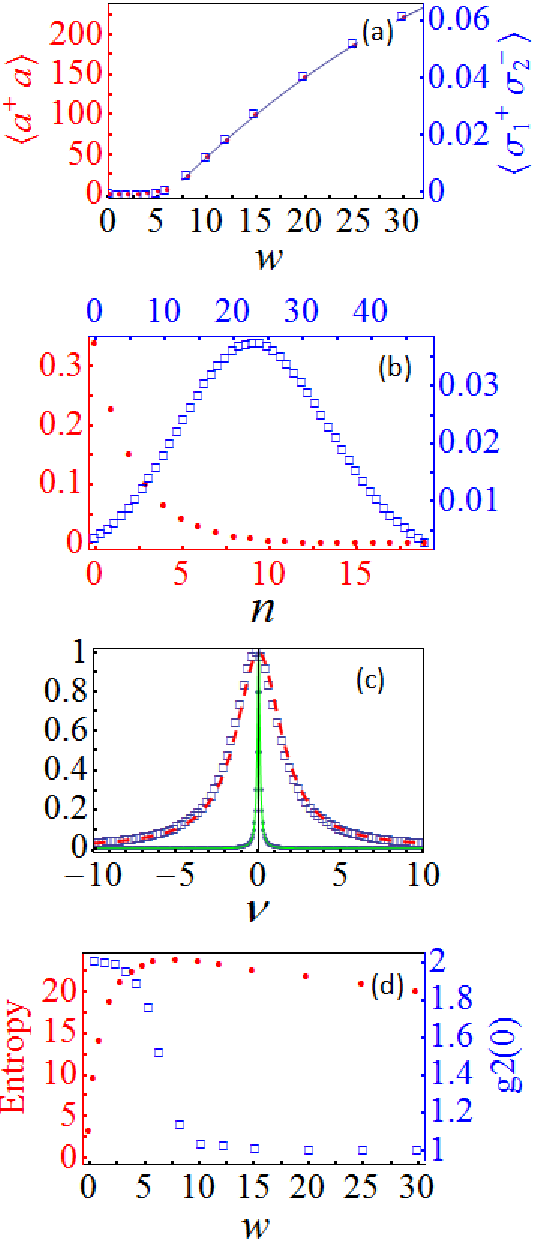}}
\caption{\label{Fig3}(color online). Calculations of laser behaviors
  described by Eq.~(\ref{eq1}) with $\Omega=1$, $\gamma=5$,
  $\kappa=1$, $1/(2T_2)=0$ and $N=30$. (a)~The average intracavity
  photon number~(red dots) and spin-spin correlation~(blue squares) as
  a function of the repumping rate $w$. The blue line is the laser
  theory prediction of the average photon
  number~\cite{Carmichael}. (b) Photon statistics of the laser below
  threshold $w=4$~(red dots) and above threshold $w=8$~(blue
  squares). (c) Normalized spectra of the laser below threshold
  $w=4$~(squares) and above threshold $w=8$~(dots). The red dashed line and green solid
  line are fitted Lorentzian lineshapes. (d) Threshold behavior
  illustrated by the intensity correlation $g^{(2)}(0)= \langle
  a^{\dagger} a^{\dagger}aa\rangle/\langle
  a^{\dagger}a\rangle^2$~(blue squares) and the entropy~(red dots) of
  the whole system as a function of the repumping rate.}
\end{figure}

Fig.~\ref{Fig3}(a) shows the average intracavity photon number of the
laser as a function of the repumping rate, where the threshold is
evident. This result confirms the conventional laser theory
prediction~\cite{Carmichael}. Interestingly, the spin-spin correlation
$\langle\sigma_j^+\sigma_k^-\rangle$ above the threshold is directly
proportional to the photon number, which shows that the collective
photon emission plays an essential role for the laser.  In
Fig.~\ref{Fig3}(b), we show that the photon statistics of the laser
changes from thermal below threshold to Poisson above threshold.  In
Fig.~\ref{Fig3}(c), we demonstrate that the laser linewidth narrows
considerably as one goes above threshold.  Finally, in
Fig.~\ref{Fig3}(d), the laser threshold behavior is characterized by
the intensity correlation $g^{(2)}(0)$ and the entropy of the whole
system. It can be seen that $g^{(2)}(0)$ jumps from two below
threshold to one above threshold with the entropy increasing and
saturating. It is remarkable to have an exact solution to this
fundamental system and to be able to rigorously confirm  standard
laser theory results. As discussed earlier, those results are typically 
based on various kinds of
analytic approximations necessary to make the problem tractable.

\begin{figure}[h]
  \centerline{\includegraphics[width=0.9\linewidth, angle=0]{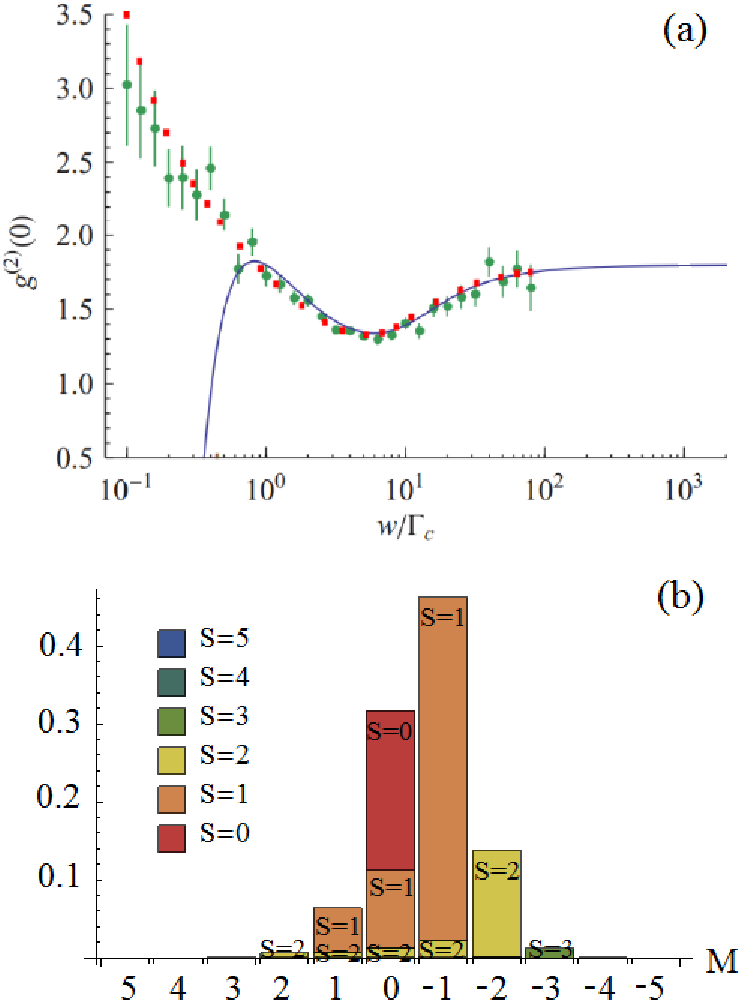}}
\caption{\label{Fig4}(color online). (a)~Comparison of the second order
  intensity correlation $g^{(2)}(0)= \langle a^{\dagger}
  a^{\dagger}aa\rangle/(a^{\dagger}a)^2$ as a function of the repump
  rate in the steady state superradiance. The green symbols show the
  Monte Carlo results including the statistical errors for $N=10$ atoms
  from Ref.~\cite{Meiser10}. The red squares show the present calculation
  using the {\em SU}(4) theory. And the blue solid line shows the
  semiclassical results from Ref.~\cite{Meiser10}. (b)~Atom
  statistics for $w=0.1\Gamma_c$. The length of the bars represents
  populations of the $|S,M\rangle$ states.}
\end{figure}
\section{Application to Steady state superradiance}
As a second example, we apply our approach to steady-state
superradiance as previously proposed~\cite{Meiser09} and demonstrated
in a recent experiment~\cite{Thompson12}. The steady-state
superradiance represents a novel regime of cavity quantum
electrodynamics, where the highly coherent collective atomic dipole
induces an extremely narrow linewidth for the generated light. The
bad-cavity mode only plays a role as the source of collective coupling
for the atoms and the definition of the spatial mode for the output
light~\cite{Meiser10}. The behavior of this system is also described
by a master equation Eq.~(\ref{eq1}), but in a completely different
parameter regime to the conventional laser. For steady-state superradiance, the
  vacuum Rabi splitting is much less than the cavity linewidth,
  $\sqrt{N}\,\Omega\ll\kappa$, and equivalently the photon number per
  atom in the cavity is much less than unity.

We present here calculations of the second order intensity correlation
$g^{(2)}(0)$ in steady-state as a function of the repump rate. As
shown in Fig.~\ref{Fig4}(a), the agreement of the present calculation
and the quantum Monte Carlo result from Ref.~\cite{Meiser10} is within
error bars. The quantum Monte Carlo simulations were significantly
more numerically intensive. In the weak pumping limit, the light
exhibits strongly super-Poissonian fluctuations and deviates
remarkably from the semiclassical prediction~(blue line in
Fig.~\ref{Fig4}(a)). The failure of the semiclassical prediction in
the weak pumping limit indicates that the atoms are in a
highly-correlated state. To reveal the atomic states in this case, we
apply the techniques of projecting the density operator in the
$P_{q,q_3,\sigma_3}$ representation onto the $|S,M\rangle$
representation and obtain the atomic populations. The inset of
Fig.~\ref{Fig4} shows explicitly that the atoms are mainly pumped into
long-lived collective subradiant states~\cite{Dicke54}
$|S=0,M=0\rangle$ and $|S=1,M=-1\rangle$. From $|S=0,M=0\rangle$, the
atoms can only be repumped to $|S=1,M=1\rangle$, from which they
rapidly emit two photons and relax to $|S=1,M=-1\rangle$.  Therefore,
our methods have enabled us to reveal detailed information about the
underlying quantum dynamics.
\section{conclusion}
In conclusion, we have formulated and applied a {\em SU}(4) theory to
numerically solve the quantum master equation, which has reduced the
exponential scaling of the problem to cubic in $N$. We have developed
powerful methods to transform the density operator in the {\em SU}(4)
basis representation to the $|S,M\rangle$ representation. This has
enabled us to efficiently diagonalize the whole density matrix and
thus provided complete information about the system, including state
information and functional properties of the density operator. We have
included lasing and steady-state superradiance as examples in order to
illustrate the potential for this method. The method described here
will find numerous applications for simulating open quantum systems
with large system size.
\begin{acknowledgments}
We acknowledge stimulating discussions with J. Cooper and
D. Meiser. This work has been supported by the DARPA QuASAR program
and the NSF.
\end{acknowledgments}
\appendix
\section{SU(4) Algebra}\label{app1}

 In order to see how the superoperators~[Eq.~(\ref{super})] are related to
 generators of the  {\em SU}(4) group~(Gell-Mann matrices),
consider first the fundamental one atom case. We interpret the
$2\times2$ density matrix as a $4\times1$ vector in the representing
vector space ({\em i.e.}~Liouville space).
\begin{equation}
\begin{pmatrix}
a & c  \\
d & b
\end{pmatrix}\rightarrow\begin{pmatrix}
a  \\
c \\
d \\
b
\end{pmatrix}.
\end{equation}
The relations are then given by
\begin{equation}
\begin{split}
  \mathcal{Q}_{\pm}\rightarrow\frac{1}{2}(\lambda_9\pm
  i\lambda_{10})&,\;\;\;\mathcal{Q}_3\rightarrow\frac{1}{4}\lambda_3
  +\frac{1}{4\sqrt{3}}\lambda_8+\sqrt{\frac{1}{6}}\lambda_{15}\\
  \Sigma_{\pm}\rightarrow\frac{1}{2}(\lambda_6\pm
  i\lambda_{7})&,\;\;\;\Sigma_3\rightarrow
  -\frac{1}{4}\lambda_3+\frac{\sqrt{3}}{4}\lambda_8\\
  \mathcal{M}_{\pm}\rightarrow\frac{1}{2}(\lambda_4\pm
  i\lambda_{5})&,\;\;\;\mathcal{M}_3\rightarrow\frac{1}{4}\lambda_3+\frac{\sqrt{3}}{4}\lambda_8\\
  \mathcal{N}_{\pm}\rightarrow\frac{1}{2}(\lambda_{11}\pm
  i\lambda_{12})&,\;\;\;\mathcal{N}_3\rightarrow-\frac{1}{4}\lambda_3
  +\frac{1}{4\sqrt{3}}\lambda_8+\sqrt{\frac{1}{6}}\lambda_{15}\\
  \mathcal{U}_{\pm}\rightarrow\frac{1}{2}(\lambda_1\pm
  i\lambda_{2})&,\;\;\;\mathcal{U}_3\rightarrow\frac{1}{2}\lambda_3\\
  \mathcal{V}_{\pm}\rightarrow\frac{1}{2}(\lambda_{13}\pm
  i\lambda_{14})&,\;\;\;\mathcal{V}_3\rightarrow
  -\frac{1}{2\sqrt{3}}\lambda_8+\sqrt{\frac{1}{6}}\lambda_{15}.
\end{split}
\end{equation}
The commutation relations of the superoperators are given in both
Ref.~\cite{Hartmann12} and \cite{Pfeifer03}. We can also identify six
{\em SU}(2) subalgebras,
\begin{equation} [\mathcal{O}_+,\mathcal{O}_-]=2\mathcal{O}_3,\;\;\;
  [\mathcal{O}_3,\mathcal{O}_{\pm}]=\pm\mathcal{O}_{\pm},
\end{equation}
so that it is useful to define six corresponding quadratic
superoperators
$\mathcal{O}^2=\mathcal{O}_-\mathcal{O}_++\mathcal{O}_3^2+\mathcal{O}_3$,
which commute with $\mathcal {O}_3$. The {\em SU}(4) group has 3
Casimir operators, one of which is quadratic in the generators, and
the others are of higher order. The quadratic Casimir operator
$\mathcal{C}_1$ can be expressed in terms of superoperators
\begin{equation}
\mathcal{C}_1=\sum_\mathcal{O}(\mathcal{O}_-\mathcal{O}_+
+\mathcal{O}_3)+\mathcal{U}_3^2+\frac{1}{3}(\mathcal{U}_3+2\Sigma_3)^2
+\frac{1}{6}(3\mathcal{Q}_3-2\mathcal{U}_3-\Sigma_3)^2.
\end{equation}

\section{Fully symmetrical basis for {\em SU}(4) group }\label{app2}
The fundamental representation of the {\em SU}(4) group, adapted to
serve as basis of the single-spin density matrix, is given by
$u=|1\rangle\langle1|$, $d=|0\rangle\langle0|$,
$s=|1\rangle\langle0|$, $c=|0\rangle\langle1|$.  Higher order
representations can then be obtained from the fundamental
representation and the symmetry type, which is described by the Young
Tableau.

The basis for the fully symmetrical case is defined as
\begin{equation}
  P_{q,q_3,\sigma_3}=\mathcal{S}(u^\alpha d^\beta
  s^\gamma c^\delta),
\end{equation}
which are eigenstates of both $\mathcal{O}^2$ and
$\mathcal{O}_3$~\cite{Hartmann12}, with eigenvalues
\begin{equation}
\mathcal{O}^2P_{q,q_3,\sigma_3}^{(\mathrm{s})}=o(o+1)P_{q,q_3,\sigma_3}^{(\mathrm{s})},\;\;\;
\mathcal{O}_3P_{q,q_3,\sigma_3}^{(\mathrm{s})}=o_3P_{q,q_3,\sigma_3}^{(\mathrm{s})},
\end{equation}
where $o\in\{q,\sigma,m,n,u,v\}$. The eigenvalues are not independent,
but can be expressed in terms of $\alpha,\gamma,\beta,\delta$:
\begin{equation}
\begin{array}{cc}
q=(\alpha+\beta)/2,&q_3=(\alpha-\beta)/2,\\
\sigma=(\gamma+\delta)/2,&\sigma_3=(\gamma-\delta)/2,\\
m=(\alpha+\delta)/2,&m_3=(\alpha-\delta)/2;\\
n=(\gamma+\beta)/2,&n_3=(\gamma-\beta)/2;\\
u=(\alpha+\gamma)/2,&u_3=(\alpha-\gamma)/2;\\
v=(\delta+\beta)/2,&v_3=(\delta-\beta)/2.
\end{array}
\end{equation}
Then it is straightforward to determine actions of all the raising and
lowering superoperators on $P_{q,q_3,\sigma_3}$,
\begin{equation}\label{act}
\begin{split}
\mathcal{Q}_{\pm}P_{q,q_3,\sigma_3}&=(q\mp q_3)P_{q,q_3\pm 1,\sigma_3},\\
\Sigma_{\pm}P_{q,q_3,\sigma_3}&=(\sigma\mp \sigma_3)P_{q,q_3,\sigma_3\pm 1},\\
\mathcal{M}_{\pm}P_{q,q_3,\sigma_3}&=(m\mp m_3)P_{q\pm 1/2,q_3\pm 1/2,\sigma_3 \pm 1/2},\\
\mathcal{N}_{\pm}P_{q,q_3,\sigma_3}&=(n\mp n_3)P_{q\mp 1/2,q_3\pm 1/2,\sigma_3 \pm 1/2},\\
\mathcal{U}_{\pm}P_{q,q_3,\sigma_3}&=(u\mp u_3)P_{q\pm 1/2,q_3\pm 1/2,\sigma_3 \mp 1/2},\\
\mathcal{V}_{\pm}P_{q,q_3,\sigma_3}&=(v\mp v_3)P_{q\mp 1/2,q_3\pm 1/2,\sigma_3 \mp 1/2}.
\end{split}
\end{equation}
We note that the fully symmetrical basis are also eigenstates of the
quadratic Casimir operator $\mathcal{C}_1$ with common eigenvalue
$3N(N+4)/8$.

Analogous actions for the photon part are the simple harmonic
oscillator relations:
\begin{eqnarray}
  a\,\bigl|n\bigr>&=&\sqrt{n}\,\bigl|n-1\bigr>\,,\nonumber\\
  a^{\dag}\,\bigl|n\bigr>&=&\sqrt{n+1}\,\bigl|n+1\bigr>\,.
\end{eqnarray}

\section{$|S,M\rangle\langle S,M'|$ representation}\label{app3}
In order to project the density operator from the
{\em SU}(4) basis onto the $|S,M\rangle\langle S,M'|$ representation
, let us first show that $M$ and $M'$ are related to the $P_{q,q_3,\sigma_3}^{(\mathrm{s})}$ by $q_3+\sigma_3=M$
 and $q_3-\sigma_3=M'$.
To see this, defining $\hat{J}_3=\sum_{j=1}^N\sigma_j^{(3)}/2$, we could get
\begin{equation}
\begin{split}
  \hat{J}_3P_{q,q_3,\sigma_3}^{(\mathrm{s})}&=\frac{1}{2}(\alpha+\gamma-
  \beta-\delta)P_{q,q_3,\sigma_3}^{(\mathrm{s})}=(q_3+\sigma_3)P_{q,q_3,\sigma_3}^{(\mathrm{s})},\\
  P_{q,q_3,\sigma_3}^{(\mathrm{s})}\hat{J}_3&=\frac{1}{2}(\alpha+\delta-
  \beta-\gamma)P_{q,q_3,\sigma_3}^{(\mathrm{s})}=(q_3-\sigma_3)P_{q,q_3,\sigma_3}^{(\mathrm{s})},
\end{split}
\end{equation}
and by definition, we have
\begin{equation}
\begin{split}
\hat{J}_3|S,M\rangle\langle S,M'|&=M|S,M\rangle\langle S,M'|,\\
|S,M\rangle\langle S,M'|\hat{J}_3&=M'|S,M\rangle\langle S,M'|.
\end{split}
\end{equation}
Therefore, the complex coefficients from the  $P_{q,q_3,\sigma_3}^{(\mathrm{s})}$
 basis contributing to the matrix element for
$|S,M\rangle\langle S,M'|$ all satisfy $q_3+\sigma_3=M$ and
$q_3-\sigma_3=M'$.

With this in mind, we now describe a systematic algorithm to obtain
the density matrix elements $D_{S,M,M'}^{m,n}$ from the {\em SU}(4)
expansion coefficients $C_{q,q_3,\sigma_3}^{m,n}$.  We illustrate our
method by considering in detail the elementary case of three atoms.
The density matrix in the $|S,M\rangle\langle S,M'|$ representation is
block diagonal in $S$; the block matrices for all $S$ can be arranged
in the shape of a pyramid as shown in Fig.~1(c).  For instance, the
base layer corresponds to $S=N/2$, with the matrix dimension being
$(N+1)^2$. The second layer has $S=N/2-1$ and dimension $(N-1)^2$, and
so on.  Furthermore there are $n_S$ copies associated with each layer,
so that $\sum_{S}(2S+1)n_S=2^N$. Taking $N=3$ for example, there are
two layers, $S=3/2$ and $S=1/2$ with $n_{3/2}=1$ and $n_{1/2}=2$, so
that the Hilbert space dimension is $(3+1)+2\times(1+1)=2^3$.

The density matrix needs to be built from the bottom layer upwards. In
the bottom layer, we find that the only element contributing to
$|N/2,N/2\rangle\langle N/2,N/2|$ is
$P_{N/2,N/2,0}^{(\mathrm{s})}$. So the top left corner is
$D_{N/2,N/2,N/2}^{m,n}=C_{N/2,N/2,0}^{m,n}$. We next apply the
lowering operator $\hat{J}_-=\sum_{j=1}^N\sigma_j^-$ to iteratively
generate $D_{N/2,N/2,M}^{m,n}$, with $M=N/2-1,\ldots,-N/2$.  To do
this, we need the recursion relation
\begin{equation}\label{rec}
\begin{split}
D_{S,M,M'-1}^{m,n}&=\langle S,M|\rho^{m,n}|S,M'-1\rangle=
\frac{\langle S,M|\rho^{m,n}\hat{J}_-|S,M'\rangle}{\sqrt{(S+M')(S-M'+1)}}\\
&=\frac{\langle S,M|(\mathcal{U}_++\mathcal{V}_+)\rho^{m,n}|S,M'\rangle}{\sqrt{(S+M')(S-M'+1)}}.
\end{split}
\end{equation}
Therefore, with the actions of the raising and lowering
operators~[Eq.~(\ref{act})], we can derive all $D_{N/2,N/2,M'}^{m,n}$,
{\em i.e.}~the first row of the bottom layer.  Using the fact that the
density matrix is Hermitian and
$C_{q,q_3,\sigma_3}^{m,n}=(C_{q,q3,-\sigma_3}^{m,n})^*$, we could get
all the elements for the first column by
$D_{N/2,M',N/2}^{m,n}=(D_{N/2,N/2,M'}^{m,n})^*$.  By repeatedly
applying the recursion relation~[Eq.~(\ref{rec})] in each row, we then
construct the full base layer.  As an explicit example, we have
constructed the bottom layer, {\em i.e.}~$S=3/2$ for the three atom
case,
\begin{equation}\label{ma}
  \bordermatrix{
    &\langle\frac32,\frac32|&\langle\frac32,\frac12|&\langle\frac32,-\frac12|&\langle\frac32,-\frac32|\cr
    |\frac32,\frac32\rangle&C_{3/2, 3/2, 0}^{m,n}&\frac{C_{1, 1, 1/2}^{m,n}}{\sqrt{3}}
    &\frac{C_{1/2, 1/2, 1}^{m,n}}{\sqrt{3}}&C_{0, 0, 3/2}^{m,n}\cr
    |\frac32,\frac12\rangle&\frac{C_{1, 1, -1/2}^{m,n} }{\sqrt{3}}
    &\frac{C_{3/2, 1/2, 0}^{m,n}+C_{1/2, 1/2, 0}^{m,n}}{3}
    &\frac{C_{1, 0, 1/2}^{m,n}+C_{0, 0, 1/2}^{m,n}}{3}&\frac{C_{1/2, -1/2, 1}^{m,n}}{\sqrt{3}}\cr
    |\frac32,-\frac12\rangle&\frac{C_{1/2, 1/2, -1}^{m,n}}{\sqrt{3}}&
    \frac{C_{1, 0, -1/2}^{m,n}+C_{0, 0, -1/2}^{m,n}}{3}
    &\frac{C_{3/2, -1/2, 0}^{m,n}+C_{1/2, -1/2, 0}^{m,n}}{3}&\frac{C_{1, -1, 1/2}^{m,n}}
      {\sqrt{3}}   \cr
    |\frac32,-\frac32\rangle&C_{0, 0, -3/2}^{m,n}&\frac{C_{1/2, -1/2, -1}^{m,n}}{\sqrt{3}}
     &\frac{C_{1, -1, -1/2}^{m,n}}{\sqrt{3}}
     &C_{3/2, -3/2, 0}^{m,n}
  }.
\end{equation}
In order to illustrate the use of the recursion relation, we now show
how to get $D_{3/2,1/2,1/2}^{m,n}$ from
$D_{3/2,1/2,3/2}^{m,n}$. Because
$\mathcal{V}_+P_{3/2,1/2,0}^{(\mathrm{s})}=
P_{1,1,-1/2}^{(\mathrm{s})}$ and
$\mathcal{U}_+P_{1/2,1/2,0}^{(\mathrm{s})}=
P_{1,1,-1/2}^{(\mathrm{s})}$ , we have $D_{3/2,1/2,1/2}^{m,n}=(C_{3/2,
  1/2, 0}^{m,n}+C_{1/2, 1/2, 0}^{m,n})/\sqrt{3}/ \sqrt{3}$.

To construct the next layer, we thus find out the top left matrix
element first, and then apply the same procedure as before to
determine the rest of the matrix elements. Let us first examine the
three atom case. The $S=1/2$ layer has two copies, each of which is a
$2\times 2$ matrix. To find the top left element
$D_{1/2,1/2,1/2}^{m,n}$, noticing the constraint imposed by the trace
of the density matrix, we derive
$2D_{1/2,1/2,1/2}^{m,n}+D_{3/2,1/2,1/2}^{m,n}=C_{3/2, 1/2, 0}^{m,n}$
so that $D_{1/2,1/2,1/2}^{m,n}=(2C_{3/2, 1/2, 0}^{m,n}-C_{1/2, 1/2,
  0}^{m,n})/6$. By applying the same method as in the bottom layer, we
construct the block matrix for $S=1/2$ layer
\begin{equation}\label{ma1}
  \bordermatrix{
    &\langle\frac12,\frac12|&\langle\frac12,-\frac12|\cr%
    |\frac12,\frac12\rangle&\frac{2C_{3/2, 1/2, 0}^{m,n}-C_{1/2, 1/2, 0}^{m,n}}{6}&
    \frac{C_{1, 0, 1/2}^{m,n}-2C_{0, 0, 1/2}^{m,n}}{6}\cr
    |\frac12,-\frac12\rangle&\frac{C_{1, 0, -1/2}^{m,n}-2C_{0, 0, -1/2}^{m,n}}{6}
    &\frac{2C_{3/2, -1/2, 0}^{m,n}-C_{1/2, -1/2, 0}^{m,n}}{6}\cr
    }.
\end{equation}

Therefore in general, if we suppose that we have constructed the block
matrix for $S'>S$, the formula to find the top left matrix element
$D_{S,S,S}^{m,n}$ for layer $S$ is
\begin{equation}
\sum_{S\leq S'\leq N/2}n_{S'}D_{S',S,S}^{m,n}=C_{N/2,S,0}^{m,n}.
\end{equation}
Having the top left matrix element for each layer $S$, we can easily
construct the $(2S+1)\times(2S+1)$ block matrix by applying the
recursion relation based on the angular momentum lowering
operator. Repeated iteration of these steps systematically fills in
all sites of the pyramid.

\end{document}